\documentclass[twocolumn,preprintnumbers,prd,nofootinbib]{revtex4-1}

\usepackage{amsmath,amssymb,slashed}
\usepackage{hyperref}
\usepackage{url}
\usepackage{breakurl}
\usepackage{graphicx}

\def\app#1#2{%
  \mathrel{%
    \setbox0=\hbox{$#1\sim$}%
    \setbox2=\hbox{%
      \rlap{\hbox{$#1\propto$}}%
      \lower1.1\ht0\box0%
    }%
    \raise0.25\ht2\box2%
  }%
}

\begin{document}

\preprint{UCB-PTH-15/15}

\title{Composite Models for the 750 GeV Diphoton Excess}

\author{Keisuke Harigaya and Yasunori Nomura}
\affiliation{Berkeley Center for Theoretical Physics, Department of Physics, 
  University of California, Berkeley, CA 94720}
\affiliation{Theoretical Physics Group, Lawrence Berkeley National 
  Laboratory, Berkeley, CA 94720}

\begin{abstract}
We present composite models explaining the diphoton excess of mass around 
$750~{\rm GeV}$ recently reported by the LHC experiments.
\end{abstract}

\maketitle

\section{Introduction}
\label{sec:intro}

Recently, the existence of a diphoton excess of mass around $750~{\rm GeV}$ 
has been reported by both the ATLAS~\cite{ATLAS} and CMS~\cite{CMS} 
experiments at the LHC.  The signals are still only $3.6~\sigma$ and 
$2.6~\sigma$ in the respective experiments, but if confirmed, this would 
indicate the long awaited discovery of new physics at the TeV scale.

Because of the Landau-Yang theorem~\cite{Landau:1948kw}, a particle 
decaying into two photons must have either spin~0, 2, or higher. 
Assuming spin~0, it is natural to postulate that the particle is 
a composite state of some strong dynamics around the TeV scale, since 
it would then not introduce any new hierarchy problem beyond that 
of the standard model Higgs boson, which may have an environmental 
understanding~\cite{Agrawal:1997gf}.

In fact, the relevance of strong dynamics is also suggested by the 
data.  Suppose the diphoton resonance is a scalar field $S$ of mass 
$m_S \simeq 750~{\rm GeV}$, whose couplings to the gluon and photon 
are induced by the following interaction
\begin{equation}
  {\cal L} = \lambda S Q \bar{Q},
\label{eq:lambda}
\end{equation}
where $\lambda$ is a coupling constant, and $(Q, \bar{Q})$ is a heavy 
vector-like fermion of mass $m_Q$ charged under the standard model 
gauge group, $SU(3)_C \times SU(2)_L \times U(1)_Y$.  Assuming that 
$(Q, \bar{Q})$ is an $SU(3)_C$ triplet and has charge $q$, its loop 
generates
\begin{equation}
  {\cal L} \sim \frac{1}{24\pi^2} \frac{\lambda S}{m_Q} 
    \biggl( \frac{1}{2} g_3^2 G^{a\mu\nu} G^a_{\mu\nu} 
    + 3 (q e)^2 F^{\mu\nu} F_{\mu\nu} \biggr),
\label{eq:SGG-SFF}
\end{equation}
where $g_3$ and $e$ are the QCD and QED gauge couplings, respectively, 
and $G^a_{\mu\nu}$ and $F_{\mu\nu}$ are the corresponding field strengths. 
The production cross section of $S$ through gluon fusion times the 
branching ratio into diphoton is then given by
\begin{equation}
  \sigma_{pp \rightarrow S} B_{S \rightarrow AA} 
  \simeq 0.2~{\rm fb} \times \lambda^2 \left( \frac{q}{2/3} \right)^4 
    \left( \frac{600~{\rm GeV}}{m_Q} \right)^2.
\label{eq:sigma-B}
\end{equation}
Here, we have normalized $m_Q$ by its experimental lower 
bound~\cite{Aad:2015tba}.  (Note also that the decay of $S$ into 
$Q \bar{Q}$ must be kinematically forbidden in order not to suppress 
the diphoton signal, giving $m_Q > m_S/2 \simeq 375~{\rm GeV}$.) 
We find that to reproduce the observed excess, which requires 
$\sigma_{pp \rightarrow S} B_{S \rightarrow AA} \sim 10~{\rm fb}$, 
the coupling $\lambda$ must be rather large.  This suggests the 
existence of strong dynamics behind the physics generating couplings 
between the resonance of interest and standard model gauge bosons.

Motivated by these considerations, in this paper we present models 
in which the observed diphoton excess arises from a composite particle 
of some hidden strong gauge interactions.  We present a scenario in 
which the particle is a state around the dynamical scale (a hidden 
glueball or a hidden eta prime) as well as a scenario in which it is 
lighter (a hidden pion).  We see that some of the models have possible 
tensions with diboson searches at $8~{\rm TeV}$, but others do not. 
In particular, we find that the model having the charge assignment 
consistent with $SU(5)$ grand unification has a parameter region in 
which the observed diphoton excess is reproduced without contradicting 
the other data.  In general, the models described here yield multiple 
resonances in the TeV region, which may be observed in the future 
LHC data.

In the appendix, we present a general analysis of constraints from 
the $8~{\rm TeV}$ data in the case that the scalar resonance is coupled 
to the standard model gauge fields only through dimension-5 operators.

\section{Hidden Glueball:\ Minimal Model}
\label{sec:glue-minimal}

Consider a hidden gauge group $G_H$, with the dynamical scale (the 
mass scale of generic resonances) $\Lambda$.  For simplicity, we take 
$G_H$ to be $SU(N)$.  We also introduce a vector-like hidden quark 
$(Q,\bar{Q})$ of mass $m$, whose charges under $G_H$ and the standard 
model $SU(3)_C$ and $U(1)_Y$ are given in Table~\ref{tab:glueball}. 
Here, $Q$ and $\bar{Q}$ represent left-handed Weyl spinors.
\begin{table}[t]
\begin{center}
\begin{tabular}{c|ccc}
           &        $G_H$ &    $SU(3)_C$ & $U(1)_Y$ \\ \hline
 $Q$       &       $\Box$ & $\bar{\Box}$ &      $a$ \\
 $\bar{Q}$ & $\bar{\Box}$ &       $\Box$ &     $-a$ \\
\end{tabular}
\end{center}
\caption{Charge assignment of a hidden ``heavy'' quark.  Here, $a \neq 0$, 
 and $Q$ and $\bar{Q}$ are left-handed Weyl spinors.}
\label{tab:glueball}
\end{table}

For $m \gtrsim \Lambda$, the hidden quark can be regarded as a ``heavy'' 
quark (with respect to $\Lambda$), and the lightest hidden hadron will 
be a glueball $s$ consisting of hidden sector gauge fields, of mass
\begin{equation}
  m_s \sim \Lambda.
\label{eq:m_s}
\end{equation}
Due to the existence of $(Q,\bar{Q})$, we expect that this state couples 
to the standard model gauge fields as
\begin{equation}
  {\cal L} \sim - \frac{\Lambda^3}{4\pi m^4} s 
    \left( \frac{g_3^2}{2} G^{a\mu\nu} G^a_{\mu\nu} 
    + \frac{9 a^2 g_1^2}{5} B^{\mu\nu} B_{\mu\nu} \right),
\label{eq:s-couplings}
\end{equation}
where $a = 1,\cdots,8$ is the $SU(3)_C$ adjoint index, and $g_3$ and 
$g_1$ are the $SU(3)_C$ and $U(1)_Y$ gauge couplings, respectively.%
\footnote{Throughout the paper, we adopt the hypercharge normalization 
 such that the standard model fermions have $(q,\bar{u},\bar{d},l,\bar{e}) 
 = (1/6,-2/3,1/3,-1/2,1)$, and $g_1$ is in the $SU(5)$ normalization.}
Here and below, we count possible factors of $4\pi$ using naive 
dimensional analysis~\cite{Manohar:1983md}.

The hidden glueball state $s$ can be produced by gluon fusion (among 
others) and decay into diphoton with the branching ratio
\begin{equation}
  B_{s \rightarrow \gamma\gamma} 
  = \frac{81 a^4}{50} \frac{\cos^4\theta_W g_1^4}{8 g_3^4} 
  \simeq 0.033\, a^4,
\label{eq:B-glue}
\end{equation}
where $\theta_W$ is the Weinberg angle.  The production cross section 
of $s$ through Eq.~(\ref{eq:s-couplings}) at $13~{\rm TeV}$ $pp$ 
collisions can be estimated (after multiplying the diphoton branching 
ratio) as
\begin{equation}
  \sigma_{pp \rightarrow s} B_{s \rightarrow \gamma\gamma} 
  \sim 10~{\rm fb} \times a^4 
    \left( \frac{3.5~{\rm TeV}}{m^4/\Lambda^3} \right)^2,
\label{eq:sigma-glue}
\end{equation}
where we have used NNPDF~3.0~\cite{Ball:2014uwa} for the parton distribution 
function.  We thus find that 
\begin{equation}
  \Lambda \sim 700~{\rm GeV},
\qquad
  m \sim 1~{\rm TeV},
\label{eq:scales-glue}
\end{equation}
give $m_s$ and $\sigma_{pp \rightarrow s} B_{s \rightarrow \gamma\gamma}$ 
roughly consistent with the excess in the $13~{\rm TeV}$ data.

The model is subject to constraints from analogous high-mass diboson 
resonance searches in the $8~{\rm TeV}$ data.  Assuming that the production 
occurs through interactions in Eq.~(\ref{eq:s-couplings}), the ratio of 
the $s$ production cross sections at $8~{\rm TeV}$ and $13~{\rm TeV}$ is
\begin{equation}
  \frac{\sigma_{pp \rightarrow s}|_{\rm 8\,TeV}}
    {\sigma_{pp \rightarrow s}|_{\rm 13\,TeV}} 
  \simeq 0.21,
\label{eq:sigma-ratio}
\end{equation}
for $m_s = 750~{\rm GeV}$.  This gives $\sigma_{pp \rightarrow s} 
B_{s \rightarrow \gamma\gamma}$ close to the upper limit from 
the $8~{\rm TeV}$ data~\cite{Aad:2015mna}.  The model also gives 
definite predictions for the relative branching ratios between 
$s \rightarrow \gamma\gamma, ZZ$, and $Z\gamma$
\begin{align}
  \frac{B_{s \rightarrow ZZ}}{B_{s \rightarrow \gamma\gamma}} 
  &= \tan^4\theta_W \simeq 0.09,
\label{eq:B_ZZ-GG}\\
  \frac{B_{s \rightarrow Z\gamma}}{B_{s \rightarrow \gamma\gamma}} 
  &= 2 \tan^2\theta_W \simeq 0.6,
\label{eq:B_ZG-GG}
\end{align}
where we have ignored the phase space factors.  With 
Eq.~(\ref{eq:sigma-ratio}), we thus obtain
\begin{equation}
  \sigma \times B|_{Z\gamma,\, {\rm 8\,TeV}} 
  \simeq 1.3~{\rm fb} \left( \frac{\sigma \times 
    B|_{\gamma\gamma,\, {\rm 13\,TeV}}}{10~{\rm fb}} \right).
\label{eq:s-Zgamma}
\end{equation}
This is consistent with the upper limit from the $8~{\rm TeV}$ 
data~\cite{Aad:2014fha}.

We note that the precise value of $\sigma \times B|_{{\rm 8\,TeV}}$ 
depends on the details of $s$ production, which we have assumed here 
to occur only through Eq.~(\ref{eq:s-couplings}).  (For an analysis of 
constraints from the $8~{\rm TeV}$ data for general dimension-5 couplings 
between $s$ and the standard model gauge fields, see the appendix.)  
The assumption of gluon fusion dominated production, however, may not 
be valid in some cases.  For example, depending on the values of $m$ 
and $\Lambda$, production of heavy resonances that are composed of 
$Q$ and $\bar{Q}$ and decay into $s$ may give comparable contributions 
to the production of single $s$ through the gluon fusion.  With 
this production mechanism, the production rates of $s$ in $8~{\rm TeV}$ 
and $13~{\rm TeV}$ $pp$ collisions will differ more, relaxing the 
constraints from the $8~{\rm TeV}$ data.  The production mechanisms 
can be differentiated experimentally, e.g., through the transverse 
momentum distribution of photons.  Detailed analyses of this issue 
are warranted.

Limits from other diboson decays of $s$, i.e.\ to $gg$ and $ZZ$, 
are weaker.

\section{Hidden Pion:\ Minimal Model}
\label{sec:pion-minimal}

We now consider a model in which the $750~{\rm GeV}$ resonance is 
a hidden ``pion,'' instead of the hidden glueball.  A virtue of this 
model is that we need to rely less on the dynamical assumption about 
the hidden sector.  As before, we take the hidden gauge group $G_H$ 
to be $SU(N)$, but now we take the hidden quarks to have charges in 
Table~\ref{tab:PNGB} and mass terms
\begin{equation}
  {\cal L} = -m_1 Q_1 \bar{Q}_1 - m_2 Q_2 \bar{Q}_2 + {\rm h.c.},
\label{eq:L_mass}
\end{equation}
where we take $m_{1,2} > 0$ without loss of generality. 
\begin{table}[t]
\begin{center}
\begin{tabular}{c|ccc|c}
             &        $G_H$ &    $SU(3)_C$ & $U(1)_Y$ & $U(1)_A$ \\ \hline
 $Q_1$       &       $\Box$ & $\bar{\Box}$ &      $a$ &    $1/3$ \\
 $Q_2$       &       $\Box$ &    ${\bf 1}$ &      $b$ &     $-1$ \\
 $\bar{Q}_1$ & $\bar{\Box}$ &       $\Box$ &     $-a$ &    $1/3$ \\
 $\bar{Q}_2$ & $\bar{\Box}$ &    ${\bf 1}$ &     $-b$ &     $-1$
\end{tabular}
\end{center}
\caption{Charge assignment of hidden ``light'' quarks.  Here, $a^2 \neq 
 b^2$, and $Q_{1,2}$ and $\bar{Q}_{1,2}$ are left-handed Weyl spinors.}
\label{tab:PNGB}
\end{table}
We assume that these masses are sufficiently smaller than the dynamical 
scale, $m_{1,2} \ll \Lambda$, so that $Q_{1,2}$ and $\bar{Q}_{1,2}$ can 
be regarded as hidden ``light'' quarks.

\subsection{Hidden Pion Dynamics}
\label{subsec:pions}

The strong $G_H$ dynamics makes the hidden quarks condensate 
\begin{equation}
  \langle Q_1 \bar{Q}_1 \rangle \approx \langle Q_2 \bar{Q}_2 \rangle 
  \equiv \langle Q \bar{Q} \rangle \approx \frac{1}{16\pi^2} \Lambda^3.
\label{eq:QQ-cond}
\end{equation}
These condensations do not break the standard model $SU(3)_C$ or 
$U(1)_Y$, since the hidden quark quantum numbers under these gauge 
groups are vector-like with respect to $G_H$~\cite{Vafa:1983tf}.  The 
spectrum below $\Lambda$ then consists of hidden pions, arising from 
spontaneous breaking of approximate $SU(4)_A$ axial flavor symmetry:
\begin{equation}
\begin{array}{ll}
  \psi \sim Q_1 \bar{Q}_1                       & ({\bf Adj}, 0), \\
  \chi \sim Q_1 \bar{Q}_2                       &    (\Box, a-b), \\
  \phi \sim Q_1 \bar{Q}_1 - Q_2 \bar{Q}_2 \quad &   ({\bf 1}, 0),
\end{array}
\label{eq:pions}
\end{equation}
where $\psi$ and $\phi$ are real scalars while $\chi$ is a complex 
scalar.  The quantum numbers in the rightmost column represent those 
under $SU(3)_C \times U(1)_Y$.  The masses of these particles are 
given by~\cite{Weinberg:1996kr}
\begin{align}
  m_\psi^2 &= 2 m_1 \frac{\langle Q \bar{Q} \rangle}{f^2} 
    + 3 \Delta_C,
\label{eq:m_psi}\\
  m_\chi^2 &= (m_1 + m_2) \frac{\langle Q \bar{Q} \rangle}{f^2} 
    + \frac{4}{3} \Delta_C + \frac{3(a-b)^2}{5} \Delta_Y,
\label{eq:m_chi}\\
  m_\phi^2 &= \frac{m_1 + 3 m_2}{2} \frac{\langle Q \bar{Q} \rangle}{f^2}.
\label{eq:m_phi}
\end{align}
Here, $f$ is the decay constant, given by
\begin{equation}
  f \approx \frac{1}{4\pi} \Lambda,
\label{eq:decay-const}
\end{equation}
and $\Delta_C$ and $\Delta_Y$ are contributions from $SU(3)_C$ and 
$U(1)_Y$ gauge loops, of order
\begin{equation}
  \Delta_C \approx \frac{g_3^2}{16\pi^2} \Lambda^2,
\qquad
  \Delta_Y \approx \frac{g_1^2}{16\pi^2} \Lambda^2.
\label{eq:Delta_CY}
\end{equation}

We assume that $\phi$ is the lightest hidden pion, which can be ensured 
by making $m_2$ smaller with respect to $m_1$.  This particle is the 
pseudo~Nambu-Goldstone boson of $U(1)_A \subset SU(4)_A$, whose charges 
are given in Table~\ref{tab:PNGB}.  The couplings of $\phi$ to the 
standard model gauge fields are determined by the $U(1)_A$-$SU(3)_C^2$ 
and $U(1)_A$-$U(1)_Y^2$ anomalies and are given by%
\footnote{The definition of our decay constant, $f$, is a factor of $2$ 
 smaller than that in Ref.~\cite{Weinberg:1996kr}:\ $f = F/2$.}
\begin{equation}
  {\cal L} = -\frac{N g_3^2}{32\sqrt{6}\pi^2 f} \phi 
    G^{a\mu\nu} \tilde{G}^a_{\mu\nu} 
  - \frac{9(a^2-b^2) N g_1^2}{80\sqrt{6}\pi^2 f} \phi 
    B^{\mu\nu} \tilde{B}_{\mu\nu},
\label{eq:pion-couplings}
\end{equation}
where $\tilde{G}^a_{\mu\nu} \equiv \epsilon_{\mu\nu\rho\sigma} 
G^{a\rho\sigma}/2$ and similarly for $\tilde{B}_{\mu\nu}$.  Therefore, 
$\phi$ can be produced by gluon fusion and decay into diphoton with 
a significant branching ratio.  Similar anomaly induced couplings 
also exist between $\psi$ and $G^{\mu\nu} \tilde{G}_{\mu\nu}$ and 
for $a \neq 0$ between $\psi$ and $G^{\mu\nu} \tilde{B}_{\mu\nu} 
= B^{\mu\nu} \tilde{G}_{\mu\nu}$, but not for other combinations 
of hidden pions and standard model gauge bosons.

\subsection{Parameter Region and Constraints}
\label{subsec:pheno}

Assuming that $\phi$ production occurs only through interactions 
in Eq.~(\ref{eq:pion-couplings}), the production cross section at 
$13~{\rm TeV}$ $pp$ collisions is given by
\begin{equation}
  \sigma_{pp \rightarrow \phi} \simeq 270~{\rm fb} 
    \left( \frac{N}{5} \frac{600~{\rm GeV}}{f} \right)^2.
\label{eq:sigma-pion}
\end{equation}
From Eq.~(\ref{eq:pion-couplings}), we find that
\begin{equation}
  B_{\phi \rightarrow \gamma\gamma} 
  = \frac{81 (a^2-b^2)^2 \cos^4\theta_W g_1^4}{50 g_3^4},
\label{eq:B-phi}
\end{equation}
so we obtain
\begin{equation}
  \sigma_{pp \rightarrow \phi} B_{\phi \rightarrow \gamma\gamma} 
  \simeq 8.9~{\rm fb} \left( \frac{N(a^2-b^2)}{5} 
    \frac{600~{\rm GeV}}{f} \right)^2.
\label{eq:sigma-phi-GG}
\end{equation}
This determines the decay constant as
\begin{equation}
  f \simeq 570~{\rm GeV}\, \frac{N(a^2-b^2)}{5} \sqrt{\frac{10~{\rm fb}}
    {\sigma_{pp \rightarrow \phi} B_{\phi \rightarrow \gamma\gamma}}}.
\label{eq:f}
\end{equation}
We then obtain from Eq.~(\ref{eq:m_phi},~\ref{eq:decay-const})
\begin{equation}
  \frac{m_1 + 3 m_2}{4} \approx \frac{m_\phi^2}{8\pi f} 
  \simeq 40~{\rm GeV} \left( \frac{570~{\rm GeV}}{f} \right),
\label{eq:m1-m2}
\end{equation}
where we have used $m_\phi = 750~{\rm GeV}$ in the last expression. 
Note, however, that this equation has an $O(1)$ uncertainty arising 
from an unknown coefficient in Eq.~(\ref{eq:decay-const}).

We thus find that the model reproduces the observed diphoton excess for
\begin{equation}
  f \sim 600~{\rm GeV},
\qquad
  m_{1,2} \sim 40~{\rm GeV}.
\label{eq:scales-phi}
\end{equation}
Since the relative branching ratios between $\phi \rightarrow 
\gamma\gamma, ZZ$, and $Z\gamma$ are the same as those of $s$ in the 
previous model, Eqs.~(\ref{eq:B_ZZ-GG},~\ref{eq:B_ZG-GG}), the present 
model is subject to the same constraints from searches of high-mass 
diboson resonances in the $8~{\rm TeV}$ data.

Other constraints on the model may come from the existence of heavier 
hidden pions, $\psi$ and $\chi$.  In particular, without higher dimension 
operators coupling the hidden and standard model sectors and suppressed 
by a scale not far above $\Lambda$, the $\chi$ particle is stable at 
collider timescales.  Once produced, it hadronizes picking up a light 
quark of the standard model.  For $|a-b| = 2/3$ and $1/3$, for example, 
this particle appears as heavy stable top and bottom scalar quarks, 
respectively.  The lower bounds on the masses of such particles are 
about $800\mbox{--}900~{\rm GeV}$~\cite{ATLAS:2014fka}.  We expect that 
$m_\chi$ satisfies this bound for $m_1 \sim m_2$ due to the contribution 
from $SU(3)_C$ gauge loop, $\Delta_C$, but if not, we can always make 
$m_\chi > 900~{\rm GeV}$ keeping $m_\phi = 750~{\rm GeV}$ by making 
$m_2$ smaller relative to $m_1$.

\subsection{Variations---Hidden Eta Prime}
\label{subsec:eta}

It is interesting to consider a parameter region of the model with 
$m_2 \ll \Lambda \ll m_1$.  In this case, we expect that the lightest 
hidden hadron is the hidden $\eta'$ particle of mass $\sim \Lambda$, 
consisting of $Q_2 \bar{Q}_2$.  The production of this particle occurs 
through that of heavy resonances involving $Q_1$ and $\bar{Q}_1$, 
which then decay into two or more hidden $\eta'$.  (The decay of 
the lightest heavy resonance into an even number of $\eta'$ would 
require parity violation, which can be provided by the $\theta$ 
term of $G_H$, $\theta_H \neq 0$.)  This makes the production cross 
sections at $8~{\rm TeV}$ and $13~{\rm TeV}$ more hierarchical than 
in Eq.~(\ref{eq:sigma-ratio}), alleviating/eliminating possible 
tensions with the $8~{\rm TeV}$ data.  The decay of the hidden 
$\eta'$ is mostly into two electroweak gauge bosons, without having 
a significant branching into two gluons.

We mention that this model may exhibit an intriguing set of signals 
if $\theta_H = 0$.  In this case, a heavy ``$\eta$'' consisting 
of $Q_1\bar{Q}_1$, which we call $\tilde{\eta}$, cannot decay into 
$\eta'$ because of parity conservation (unless the three $\eta'$ mode 
is kinematically open) and hence decays into $gg$ and $\gamma\gamma$. 
On the other hand, a heavy ``$\rho$,'' having $J^{PC} = 1^{--}$, 
decays into $\gamma$ and $\eta'$, which subsequently decays as 
$\eta' \rightarrow \gamma\gamma$.  Therefore, by choosing $m_{\eta'} 
= 750~{\rm GeV}$ and $m_{\tilde{\eta}} \simeq 1.6~{\rm TeV}$ we 
may also be able to accommodate the second, (much) weaker diphoton 
excess seen around $1.6~{\rm TeV}$ in the ATLAS data~\cite{ATLAS}. 
To be conclusive, however, a more detailed analysis is needed.

\section{Models Consistent with Unification}
\label{sec:var}

We finally discuss a model in which the charge assignment of the hidden 
quarks is consistent with $SU(5)$ grand unification~\cite{Georgi:1974sy}. 
We take the hidden quarks to be a vector-like fermion in the bifundamental 
representation of $G_H$ and $SU(5)$, as shown in Table~\ref{tab:unif}, 
and write down their mass terms as in Eq.~(\ref{eq:L_mass}).  Depending 
on the relative sizes of $\Lambda$ and $m_{1,2}$, this model realizes 
either the hidden glueball or hidden pion scenario.
\begin{table}[t]
\begin{center}
\begin{tabular}{c|cccc|c}
   & $G_H$ & $SU(3)_C$ & $SU(2)_L$ & $U(1)_Y$ & $U(1)_A$ \\ \hline
 $Q_1$       &       $\Box$ & $\bar{\Box}$ & ${\bf 1}$ &  $1/3$ &  $1/3$ \\
 $Q_2$       &       $\Box$ &    ${\bf 1}$ &    $\Box$ & $-1/2$ & $-1/2$ \\
 $\bar{Q}_1$ & $\bar{\Box}$ &       $\Box$ & ${\bf 1}$ & $-1/3$ &  $1/3$ \\
 $\bar{Q}_2$ & $\bar{\Box}$ &    ${\bf 1}$ &    $\Box$ &  $1/2$ & $-1/2$
\end{tabular}
\end{center}
\caption{Charge assignment of hidden quarks consistent with grand 
 unification.}
\label{tab:unif}
\end{table}

\subsection{Hidden Glueball}
\label{subsec:glueball_unif}

For $m_{1,2} \gtrsim \Lambda$, the lightest hidden hadron is the hidden 
glueball $s$, which interacts with the standard model gauge fields as
\begin{align}
  {\cal L} \sim & - \frac{\Lambda^3}{4\pi m_1^4} s 
      \left( \frac{g_3^2}{2} G^{a\mu\nu} G^a_{\mu\nu} 
      + \frac{g_1^2}{5} B^{\mu\nu} B_{\mu\nu} \right) 
\nonumber\\
  & - \frac{\Lambda^3}{4\pi m_2^4} s 
      \left( \frac{g_2^2}{2} W^{b\mu\nu} W^b_{\mu\nu} 
      + \frac{3 g_1^2}{10} B^{\mu\nu} B_{\mu\nu} \right),
\label{eq:s-couplings_unif}
\end{align}
where $b = 1,2,3$ is the $SU(2)_L$ adjoint index, and $g_2$ is the 
$SU(2)_L$ gauge coupling.  For $m_1 \ll m_2$, the phenomenology is 
essentially the same as the non-unified model before.  For $m_2 \ll m_1$, 
on the other hand, the production of $s$ occurs through that of heavy 
resonances involving $Q_1$ and $\bar{Q}_1$.  (In order for the lightest 
heavy resonance to decay into two $s$, we need to break parity by $\theta_H 
\neq 0$.)  The produced $s$ decays mostly into electroweak gauge bosons, 
with the relative branching ratios
\begin{equation}
  R^s_{WW} \simeq 9, \qquad
  R^s_{ZZ} \simeq 10, \qquad
  R^s_{Z\gamma} \simeq 0.7,
\label{eq:B-s_unif}
\end{equation}
where $R^s_{XY} \equiv B_{s \rightarrow XY}/B_{s \rightarrow \gamma\gamma}$. 
By choosing $m_s = 750~{\rm GeV}$, we can explain the observed diphoton 
excess.

\subsection{Hidden Pion}
\label{subsec:pion_unif}

For $m_{1,2} \ll \Lambda$, the low energy spectrum consists of hidden 
pions with the following standard model quantum numbers:
\begin{equation}
\begin{array}{ll}
  \psi({\bf Adj}, {\bf 1}, 0),\qquad &
  \chi(\Box, \Box, -5/6), \\
  \varphi({\bf 1}, {\bf Adj}, 0),\qquad &
  \phi({\bf 1}, {\bf 1}, 0),
\end{array}
\label{eq:pions_unif}
\end{equation}
where $\psi$, $\varphi$, and $\phi$ are real scalars while $\chi$ is 
a complex scalar.  In this case, we can make $\phi$ the lightest hidden 
pion and identify it with the observed $750~{\rm GeV}$ resonance. 
The production and decay of this particle are both controlled by the 
anomaly induced couplings
\begin{align}
  {\cal L} = &-\frac{N g_3^2}{16\sqrt{15}\pi^2 f} \phi 
    G^{a\mu\nu} \tilde{G}^a_{\mu\nu} 
  + \frac{3 N g_2^2}{32\sqrt{15}\pi^2 f} \phi 
    W^{b\mu\nu} \tilde{W}^b_{\mu\nu}
\nonumber\\
  &+ \frac{N g_1^2}{32\sqrt{15}\pi^2 f} \phi 
    B^{\mu\nu} \tilde{B}_{\mu\nu},
\label{eq:phi-couplings_unif}
\end{align}
giving the relation in Eq.~(\ref{eq:sigma-ratio}) and
\begin{equation}
  \frac{B_{\phi \rightarrow Z\gamma}}{B_{\phi \rightarrow \gamma\gamma}} 
  = 2 \left( \frac{9 - 5 \tan^2\theta_W}{14 \tan\theta_W} \right)^2 
  \simeq 2.
\label{eq:phi_ZG-GG_unif}
\end{equation}
This leads to
\begin{equation}
  \sigma \times B|_{Z\gamma,\, {\rm 8\,TeV}} 
  \simeq 4.0~{\rm fb} \left( \frac{\sigma \times 
    B|_{\gamma\gamma,\, {\rm 13\,TeV}}}{10~{\rm fb}} \right),
\label{eq:phi-Zgamma}
\end{equation}
yielding a possible tension with the $8~{\rm TeV}$ data~\cite{Aad:2014fha}.

Another interesting region is the one in which $\varphi$ is the lightest, 
or only, hidden pion.  A simple realization of this scenario occurs if
\begin{equation}
  m_2 < \Lambda \lesssim m_1.
\label{eq:scales_unif}
\end{equation}
In this case, the spectrum below $\Lambda$ consists only of
\begin{equation}
  \varphi \sim Q_2 \bar{Q}_2 \quad ({\bf 1}, {\bf Adj}, 0),
\label{eq:varphi}
\end{equation}
of mass
\begin{equation}
  m_\varphi^2 = 2m_2 \frac{\langle Q \bar{Q} \rangle}{f^2} 
    + 2 \Delta_L,
\label{eq:m_varphi}
\end{equation}
where $\Delta_L \approx (g_2^2/16\pi^2)\Lambda^2$ represents the 
contribution from $SU(2)_L$ gauge loop.  The only anomaly induced coupling 
between this particle and the standard model gauge fields is
\begin{equation}
  {\cal L} = \frac{3 N g_2 g_1}{16\sqrt{15}\pi^2 f} 
    \varphi^b W^{b\mu\nu} \tilde{B}_{\mu\nu},
\label{eq:varphi-couplings-unif}
\end{equation}
so its production must occur through that of heavy resonances involving 
$Q_1$ and $\bar{Q}_1$ which decay into two or more $\varphi$ (with two 
$\varphi$ from the lightest heavy resonance needing $\theta_H \neq 0$). 
This, therefore, does not lead to the (potentially problematic) relation 
in Eq.~(\ref{eq:sigma-ratio}).

The probability of producing each of the three isospin components of 
$\varphi$---$\varphi^0$ and $\varphi^\pm$---is $1/3$.  The produced 
$\varphi^0$ decays into $ZZ$, $\gamma\gamma$, and $Z\gamma$ with the 
branching ratios
\begin{align}
  & B_{\varphi^0 \rightarrow ZZ} = B_{\varphi^0 \rightarrow \gamma\gamma} 
  = 2 \sin^2\theta_W \cos^2\theta_W 
  \simeq 0.35,
\nonumber\\
  & B_{\varphi^0 \rightarrow Z\gamma} 
  = \bigl( \cos^2\theta_W - \sin^2\theta_W \bigr)^2 
  \simeq 0.30,
\end{align}
while $\varphi^\pm$ into $WZ$ and $W\gamma$ with $\sin^2\theta_W 
\simeq 0.23$ and $\cos^2\theta_W \simeq 0.77$, respectively.  Thus, 
for each production of the heavy resonance, the probabilities of 
having $\gamma\gamma$ and $Z\gamma$ in the final state are
\begin{align}
  & P_{\gamma\gamma} = \frac{2}{3} B_{\varphi^0 \rightarrow \gamma\gamma} 
    - \frac{1}{9} B_{\varphi^0 \rightarrow \gamma\gamma}^2 
  \simeq 0.22,
\label{eq:P_GG}\\
  & P_{Z\gamma} = \frac{2}{3} B_{\varphi^0 \rightarrow Z\gamma} 
    - \frac{1}{9} B_{\varphi^0 \rightarrow Z\gamma}^2 
  \simeq 0.19,
\label{eq:P_ZG}
\end{align}
where we have assumed that each heavy resonance leads to two $\varphi$. 
This implies that the production cross section of heavy resonances 
of order
\begin{equation}
  \sigma \sim \frac{10~{\rm fb}}{P_{\gamma\gamma}} \simeq 45~{\rm fb},
\label{eq:sigma-needed}
\end{equation}
allows for explaining the diphoton excess.

Estimating the production cross section of heavy resonances by 
that of a heavy vector-like quark of mass $\sim {\rm TeV}$, we 
obtain~\cite{Aguilar-Saavedra:2013qpa}
\begin{equation}
  \sigma \sim 60~{\rm fb} \left( \frac{N}{2} \right).
\label{eq:sigma}
\end{equation}
We thus expect that the model yields the observed level of the diphoton 
excess for
\begin{equation}
  m_1 \sim 1~{\rm TeV},
\qquad
  N \sim O(1).
\label{eq:Lambda}
\end{equation}
Indeed, with this choice of $m_1$, the ratio of the production cross 
sections at $8$ and $13~{\rm TeV}$ is
\begin{equation}
  \frac{\sigma|_{\rm 8\,TeV}}{\sigma|_{\rm 13\,TeV}} 
  \sim 0.05\mbox{--}0.1,
\label{eq:sigma-ratio_unif}
\end{equation}
which is about a factor of $2$ smaller than in Eq.~(\ref{eq:sigma-ratio}). 
Together with
\begin{equation}
  \frac{P_{Z\gamma}}{P_{\gamma\gamma}} \simeq 0.85,
\label{eq:PZG-PGG}
\end{equation}
we find that the upper limit on the $Z\gamma$ mode from the $8~{\rm TeV}$ 
search can be safely evaded.  The limits from other diboson modes, $WZ$ 
and $W\gamma$ from $\varphi^\pm$ and $ZZ$ and $\gamma\gamma$ from 
$\varphi^0$, do not constrain the model further.

Taking $m_1 \sim 1~{\rm TeV}$ also allows for avoiding 
bounds~\cite{ATLAS:2014fka} from possible particles which are 
stable at collider timescales, e.g.\ states analogous to $\chi$ 
in Eq.~(\ref{eq:pions_unif}).  The mass of the $Q_2$ hidden quark 
is determined by the condition $m_\varphi = 750~{\rm GeV}$ as
\begin{equation}
  m_2 \sim 300~{\rm GeV}\, \frac{1~{\rm TeV}}{\Lambda},
\label{eq:m_2}
\end{equation}
through Eq.~(\ref{eq:m_varphi}).  We thus find that the model reproduces 
the observed diphoton excess while being consistent with the other 
experimental data, with the choice of parameters
\begin{equation}
  m_1 \sim \Lambda \sim 1~{\rm TeV},
\qquad
  m_2 \sim 300~{\rm GeV},
\label{eq:parameters_unif}
\end{equation}
for $N \sim O(1)$.  Note that the value of $\Lambda$ can be somewhat 
smaller if it is compensated by the correspondingly larger value of 
$m_2$ to keep $m_\varphi = 750~{\rm GeV}$, see Eq.~(\ref{eq:m_2}).

\section{Discussion}
\label{sec:discuss}

In this paper, we have presented models in which the reported diphoton 
excess of mass around $750~{\rm GeV}$ is explained by composite states 
of hidden strong gauge interactions.  The models are technically 
natural---i.e.\ do not introduce any new hierarchy problem beyond 
that of the standard model Higgs boson---and are consistent with the 
other experimental data (although some of them may have potential 
tensions with the $8~{\rm TeV}$ data).  In particular, we have 
constructed a simple model consistent with $SU(5)$ grand unification, 
in which the hidden quarks are in the bifundamental representation 
of $G_H$ and $SU(5)$.  This model is consistent with the data and 
preserves gauge coupling unification at the level of the standard 
model (which is significant; see discussion in Ref.~\cite{Hall:2009nd}, 
for example).  Interestingly, some of the parameter regions have the 
ratio of the hidden quark masses, $m_1/m_2 \sim 3$, which is roughly 
consistent with the effect of renormalization group evolution between 
the TeV and unification scales.

Models presented here require a coincidence of scales between the 
dynamical scale, $\Lambda$, and the masses of hidden quarks, $m$. 
This can be explained, for example, if the $G_H$ gauge theory is in 
the strongly coupled conformal window above TeV and deviates from 
it in the infrared due to the effect of the hidden quark masses 
that originate from a common scale.  The conformal nature of the 
dynamics may also help alleviating potential cosmological problems 
associated with stable particles in the $G_H$ sector by enhancing 
their couplings to the standard model particles.

It is interesting that a simple structure described here---new $G_H$ 
gauge interactions with vector-like matter charged under both $G_H$ 
and the standard model gauge groups---explains the observed excess(es) 
while still consistent with the other experimental data.  If the theory 
presented here is true, then the LHC Run~2 would see a plethora of new 
phenomena arising from the new strong gauge interactions.

\vspace{0.1cm}
{\it Note added:}
After this paper was submitted to arXiv, we became aware of the works 
in Ref.~\cite{Kilic:2009mi} which discuss similar theories.  These 
works focus mostly on phenomenology of spin-$1$ resonant production 
of hidden hadron pairs.

\section*{Acknowledgments}

We would like to thank Lawrence Hall, Tongyan Lin, Hou Keong Lou, Michele 
Papucci, Alex Pomarol, Surjeet Rajendran, and Kathryn Zurek for discussions. 
This work was supported in part by the Director, Office of Science, 
Office of High Energy and Nuclear Physics, of the U.S.\ Department 
of Energy under Contract DE-AC02-05CH11231, by the National Science 
Foundation under grants PHY-1316783 and PHY-1521446, and by MEXT 
KAKENHI Grant Number 15H05895.

\vspace{0.1cm}
\appendix

\section{General Dimension-5 Operators}
\label{sec:appendix}

\begin{figure}[t]
\centering
  \includegraphics[width=0.9\linewidth]{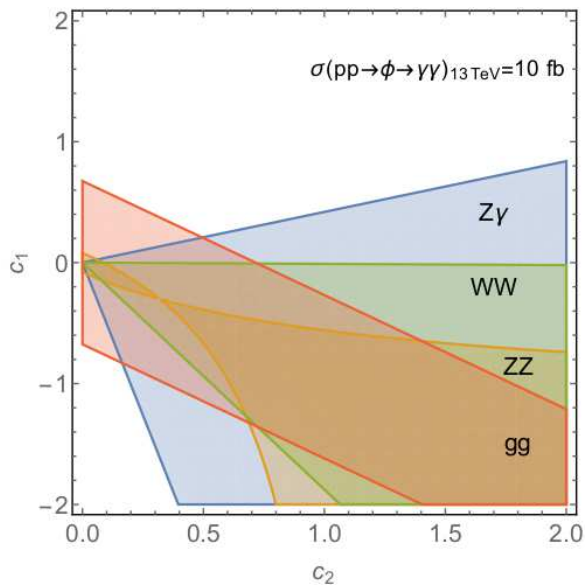}
\caption{Constraints on the coefficients of the dimension-5 operators, 
 $c_2$ and $c_1$.  The shaded regions are excluded by the searches in 
 the $8~{\rm TeV}$ data.}
\label{fig:general}
\end{figure}
In this appendix, we analyze consistency between the diphoton signal at 
$13~{\rm TeV}$ and diboson searches at $8~{\rm TeV}$, assuming general 
dimension-5 operators coupling a singlet scalar $\phi$ with the standard 
model gauge fields:
\begin{equation}
  {\cal L} = \frac{\phi}{4\pi \Lambda} 
    \biggl( g_3^2 G^{a \mu\nu} \tilde{G}^a_{\mu\nu} 
    + c_2 g_2^2 W^{b \mu\nu} \tilde{W}^b_{\mu\nu} 
    + \frac{3 c_1}{5} g_1^2 B^{\mu\nu} \tilde{B}_{\mu\nu} \biggr).
\label{eq:app-ops}
\end{equation}
(Our analysis applies if we replace $\tilde{G}^a_{\mu\nu}$ with 
$G^a_{\mu\nu}$, and similarly for $\tilde{W}^b_{\mu\nu}$ and 
$\tilde{B}_{\mu\nu}$.)  We assume that the scalar $\phi$ is produced 
via gluon fusion and decays into standard model gauge bosons. 

The constraint on each decay mode at the $8~{\rm TeV}$ LHC is given by%
\footnote{To obtain the constraint from the $gg$ channel, we assume that 
 the acceptance of the signal is $0.6$, which is the case for an octet 
 scalar decaying to two gluons.}
\begin{equation}
\begin{array}{ll}
  \sigma_{pp \rightarrow \phi} B_{\phi \rightarrow \gamma\gamma} 
    < 2~{\rm fb} \quad    & \mbox{\cite{Aad:2015mna}} \\
  \sigma_{pp \rightarrow \phi} B_{\phi \rightarrow Z\gamma} 
    < 5~{\rm fb} \quad  & \mbox{\cite{Aad:2014fha}} \\
  \sigma_{pp \rightarrow \phi} B_{\phi \rightarrow ZZ} 
    < 30~{\rm fb} \quad   & \mbox{\cite{ATLAS_comb}} \\
  \sigma_{pp \rightarrow \phi} B_{\phi \rightarrow W^+ W^-} 
    < 80~{\rm fb} \quad   & \mbox{\cite{ATLAS_comb}} \\
  \sigma_{pp \rightarrow \phi} B_{\phi \rightarrow gg} 
    < 5000~{\rm fb} \quad & \mbox{\cite{Aad:2014aqa}}
\end{array}
\end{equation}
In Fig~\ref{fig:general}, we show the constraints on $c_2$ and $c_1$ 
from the $8~{\rm TeV}$ data, choosing $\Lambda$ such that the diphoton 
signal of $10~{\rm fb}$ is obtained at $13~{\rm TeV}$.  We find that 
the limits from the $Z\gamma$ and $ZZ$ modes are constraining.  However, 
there is still a parameter region in which the observed diphoton excess 
at $13~{\rm TeV}$ is consistent with the limits from the $8~{\rm TeV}$ 
searches.

\end{document}